\documentclass{ws-p8-50x6-00}
\newcommand{\epsp}{\ensuremath{\varepsilon^\prime_K}}
\newcommand{\eps}{\ensuremath{\varepsilon_K}}

\newcommand{\ba}{\begin{eqnarray}}
\newcommand{\ea}{\end{eqnarray}}

\begin{document}

\title{ \epsp/\eps\ IN THE CHIRAL LIMIT USING LARGE $N_c$}
%Talk presented at Chiral Dynamics 2000, Jefferson Lab, Newport News,
%Virginia, USA, July 17-22, 2000.

\author{Johan Bijnens}

\address{Department of Theoretical Physics 2, Lund University\\
S\"olvegatan 14A, S 22362 Lund, Sweden}

\author{Joaquim Prades}

\address{Departamento de F\'\i sica Te\'orica y del Cosmos,
Universidad de Granada,\\
Campus de Fuente Nueva, E-18002 Granada, Spain}

\maketitle

\abstracts{
The $K\to\pi\pi$ system is analyzed in the chiral limit
and at NLO in $1/N_c$.
The $\Delta I=1/2$ rule is reproduced
and we obtain a
large value for $\varepsilon_K^\prime/\varepsilon_K$.}

\section{Introduction}

We present our calculation of \epsp/\eps\ in
the chiral limit.\cite{BP1} 
Its experimental observation
is the first new major CP-violation
result since the original discovery.\cite{Cenci}

This calculation uses the large $N_c$
approach pioneered by Bardeen et al.\cite{Bardeen}.
We  used this method previously to calculate $B_K$\cite{BP2},
the $\Delta I=1/2$-rule\cite{BP3}, the muon $g-2$
and electromagnetic mass differences.\cite{BPP}
Similar recent work is \cite{Hambye,LMD}

\section{Short-distance and $X$-boson or fictitious gauge boson scheme}

$\Delta S=1$ weak decays are produced by the
exchange of $W$-bosons. The resulting effective action, resumming
$\alpha_S\log(M_X/\nu)$ to two-loops, is now standard.
An introduction and references can be found in the
lectures by Buras.\cite{Buras} The resulting effective
action, $\Gamma_{\Delta S=1}\sim 
{ \sum_{i=1}^{10}}\, C_i(\nu)  \int {\rm d}^4 x \, Q_i(x)
+ {\rm h.c.} $
with  $Q_i(x)$ four-quark operators,
is dependent on the various schemes chosen in its definition.
This dependence should disappear in the matrix-elements of this effective
action which are usually calculated in a different way.

Because currents and densities can be matched more easily across theories,
we first replace the effective four-quark operator action by an equivalent one.
For instance, the effective action reproducing
$
Q_1(x)=\left[\overline s \gamma^\mu (1-\gamma_5)  d \right] \,
 \left[\overline u \gamma_\mu (1-\gamma_5)  u \right](x)
$ by the exchange of the heavy fictitious $X_1$
is 
\be
\Gamma_{X}\equiv g_1(\mu_c,\cdots) 
\int {\rm d}^4 y 
\, X_1^\mu \left\{ \left[\overline s \gamma_\mu (1-\gamma_5)  
d \right](x) +
 \left[\overline u \gamma_\mu (1-\gamma_5)  u \right](x) \right\}
\, .
\ee
The constants, like $g_1(\mu_c,\cdots)$,
are determined unambiguously from
matching conditions in the perturbative regime,\cite{BP1,BP2}
removing the scheme-dependence.

\section{Long-distance and results}
We calculate the matrix-elements of $X$-boson
exchange by integrating over its momentum $p_X$ 
in the Euclidean. The short-distance contribution,
$p^2_X > \mu^2$, is,
to NLO in $1/N_c$, calculable using perturbative QCD and factorization.
The long-distance part, small $p^2_X$, can be calculated
with Chiral Perturbation Theory, also done by\cite{Hambye}. 
This fails at fairly low $p_X^2$
and we have chosen a reasonable hadronic model, the ENJL model,
to extend the range to higher values. We find good matching
for the isospin zero amplitude $a_0$ and for all the imaginary parts.
Another improvement route is
the LMD method\cite{LMD}.

We obtain, to NLO in $1/N_c$ and in the chiral limit,\cite{BP1}
$a_0$ within errors and
\ba
&
B_{6\chi}^{(1/2)NDR}(2 \, {\rm GeV})= 2.5 \, \pm 0.4 \quad\quad
B_{8\chi}^{(3/2)NDR}(2 \, {\rm GeV})= 1.35 \, \pm 0.20
& \nonumber\\
&
\left|{\varepsilon_K'}/{\varepsilon_K}\right|
= (60 \pm 30 )\cdot 10^{-4}\,.  
&\ea
Including the main isospin breaking effect and final state interactions
gives\cite{BP1}
\ba
\left|{\varepsilon_K'}/{\varepsilon_K}\right|
&=& (34 \pm 18 )\cdot 10^{-4}
\ea
in reasonable agreement with the experimental value\cite{Cenci}
$(19.3\pm2.4)~10^{-4}$.

\section*{Acknowledgments}
Work partially supported  by NFR, Sweden,
EU TMR Network
EURODAPHNE (Contract No. ERBFMX-CT98-0169),
CICYT, Spain (Grant No. AEN-96/1672) and the
Junta de Andaluc\'{\i}a (Grant No. FQM-101).

\end{document}